\documentclass[twocolumn,aps,prd,tighten,amssymb,,amsmath]{revtex4}
\usepackage{amssymb}
\usepackage{epsfig}

\begin{document}

\title{Orbital Dynamics of Binary Boson Star Systems}

\author{ C.~Palenzuela$^{1}$, L.~Lehner$^{1}$, S.~L.~Liebling$^{2}$}

\affiliation{${}^1$ Department of Physics and
Astronomy, Louisiana State University, 202 Nicholson Hall, Baton
Rouge, Louisiana 70803-4001, USA\\
${}^2$ Department of Physics, Long Island University,
Brookville, New York 11548, USA}

\begin{abstract}
We extend our previous studies of head-on collisions of boson stars by considering
orbiting binary boson stars. We concentrate on equal mass binaries and study the dynamical behavior
of boson/boson and  boson/antiboson pairs. We examine the
gravitational wave output of these binaries and compare with
other compact binaries.
Such a comparison lets us probe the apparent simplicity observed in
gravitational waves produced by black hole binary systems. In our system of interest however,
there is an additional internal freedom which plays a significant role in the system's dynamics,
namely the phase of each star.
Our evolutions show rather simple behavior at early times, but large differences occur at late times for
the various initial configurations.
\end{abstract}

\maketitle

\section{Introduction}
\label{introduction}
Boson stars are compact solutions to Einstein's gravitational field equations coupled 
to a massive, complex scalar field. In addition to their interest as
soliton-like solutions akin to so-called Q-balls \cite{Qball},
they also serve as a simplified model
of an astrophysical compact object with which to probe non-linear regimes. 

In a recent work, we described an implementation of the Einstein equations 
that can evolve, among other possible sources, boson star configurations~\cite{POLL07}.
We studied the head-on collision of boson stars with equal mass but possibly differing in phase and
oscillation frequencies. The simulations revealed interesting effects arising from 
these possible differences. In particular, by adopting phase and frequency reflections the collisions
displayed markedly different behavior, from merger to repulsion.
In the present work we revisit these binaries in scenarios with non-vanishing angular momentum.

Our interest in the present work is twofold. On the one hand, we want to understand the
dynamics of the more complex scenario of orbiting stars.
The presence of angular momentum in this case gives rise to richer phenomenology arising
due to the orbiting behavior of the system and the peculiarities of single
boson stars with angular momentum. As illustrated by stationary, axisymmetric, rotating boson star
solutions whose angular momentum is quantized~\cite{YosEri97}, rotating boson stars have angular momentum
further constrained than their regular fluid counterparts.

On the other hand, the system also probes the dynamics of
compact objects in general relativity. In this context it is particularly interesting  
to analyze to what extent rather simple analysis can shed light on the orbiting behavior and
gravitational wave output of the system. We recall that binary black hole simulations have
revealed, among other features, that the dynamics and consequent radiation from binary black
holes can be reasonably well approximated by perturbative approaches well beyond the limited range of
validity initially assumed~\cite{Pan:2007nw,Buonanno:2006ui}. 
Such a result indicates that for binary black holes, the system undergoes a rather smooth transition 
from a quasi-adiabatic inspiral regime to the merger and ring-down without significant nonlinearities becoming
major factors. This rather simple behavior seems to be a consequence of the final object being
another black hole. When this black hole forms, it encircles a larger region which `cuts-away' portions of the spacetime
with strong curvature and dynamics and the spacetime is described by a perturbed black hole.

The binary boson stars considered in this work enable us to further examine compact binary systems 
in a different but certainly related context. For binary boson star cases, as for binary neutron star cases,
the evolution at early stages demonstrates simple quasi-adiabatic behavior. However, as the stars
approach each other, effects related to the stars' internal structure play a significant role.
Here we study binary boson stars, analyze their dynamical behavior  
and extract the gravitational wave output for different cases. 

As we discuss, significant differences arise from two effects. One effect
is the same as that discussed in the head-on case arising from the effective interaction energy inducing
significantly different gravitational potential wells when the stars come close to each other.

The other effect concerns the possible end states of the merger given the
presence of angular momentum. If the merger is to produce a single remnant star, presumably it would
yield, asymptotically, a stationary, spinning boson star. However, for the system to settle into one of the spinning boson stars found in~\cite{YosEri97}, it would have to shed angular momentum and/or energy to  be
consistent with one of the quantized states. 
There are several ``channels'' for the system to achieve this. One is through radiation of angular momentum
and energy; this, however, might require long dynamical times for the system to shed enough. In related
orbiting systems, like binary black holes, only a few percent of the initial mass/angular momentum~\cite{Bruegmann:2006at,loustowaves,pretorius,Baker:2006yw} is radiated. 
Thus, extraction of energy/angular momentum via gravitational waves from the merged object is likely a small
effect here as well. Other channels,
such as the  break-up of the merged star and black hole formation, will play significant roles in the dynamics of the system as we illustrate in this work.

For a particular configuration (i.e, masses and separation of the stars), the evolution of boson/boson
pair appears to be highly dependent on 
the value of the initial angular momentum $J_z$. This outcome contrasts with the results of our previous work~\cite{POLL07}
which demonstrated that the (unboosted) head-on collision ($J_z=0$) results in a remnant black hole
when the individual masses are large.

The paper is organized as follows. In Section \ref{equations} we briefly summarize
the formalism for the Einstein-Klein-Gordon system, referring the reader to previous work for more details~\cite{POLL07}.
Section~\ref{sec:id} describes the procedure for setting the initial data.
Section~\ref{analysis} recalls some of the analysis quantities that are going to be
used along this paper.
Section \ref{numerics} describes the numerical implementation of the governing equations
and the presentation of our results.
We conclude in Section~\ref{conclusions} with some final comments.


\section{The Einstein-Klein-Gordon system}
\label{equations}
As described in~\cite{POLL07},
the dynamics of a massive complex scalar field in a curved spacetime is
described by the following Lagrangian density (in
geometrical units $G=c=1$)~\cite{das}
\begin{eqnarray}\label{Lagrangian}
  {\cal L} = - \frac{1}{16 \pi} R
   + \frac{1}{2} \left[ g^{ab} \partial_a \bar \phi \partial_b \phi
   + V\left( \left|\phi\right|^2\right) \right]\, .
\end{eqnarray}
Here $R$ is the Ricci scalar, $g_{ab}$ is the spacetime metric,
$\phi$ is the scalar field,  $\bar \phi$ its complex conjugate,
and $V(|\phi|^2) = m^2~ |\phi|^2 $ the interaction potential with
$m$ the mass of the bosonic particle which has inverse length units. 
Throughout this paper Roman letters at the beginning of the
alphabet $a,b,c,..$ denote spacetime indices ranging from 0 to 3,
while letters near the middle $i,j,k,..$ range from 1 to 3,
denoting spatial indices.

This Lagrangian gives rise to the equations determining the
evolution of the metric (Einstein equations) and those governing
the scalar field behavior (Klein-Gordon Equations).
We employ the same implementation described in~\cite{POLL07}
to simulate our systems of interest, and the reader is referred to that
paper for the complete details. The only difference is with the initial
data described in the following section.


\section{Initial data}
\label{sec:id}
We adopt a simple prescription to define the initial data describing two boson
stars. Throughout this work this data will be obtained by a superposition of
two stars with linear momentum. In order to define such data, we adopt the 
spacetime and the scalar field described in ~\cite{POLL07} and boost each star either
with a Galilean or Lorentz boost. While this data satisfy the constraints only approximately,
the constraint violation measured on the initial data is at or below the truncation error.
Furthermore, although the runs presented here have been obtained with the Galilean boost,
similar behaviour is obtained when employing a Lorentzian one, since after some transient epoch
the dynamics of the system is, for the most part, insensitive to the details of the initial data definition.
The procedure for constructing the initial data can be schematically represented in a few steps
\begin{enumerate}
  \item compute the initial data for the single boson star $(i)$ centered at $x_i$, that can be
        described with the scalar field and the metric $\{ \phi^{(i)}, g_{ab}^{(i)} \}$. This step was
	explained in detail in \cite{POLL07}.
  \item boost both the scalar field and the spacetime (and their derivatives) by performing
        the Galilean or Lorentz coordinate transformation.
	The boosted fields will be denoted by $\{ {\hat{\phi}}^{(i)}, {\hat{g}}_{ab}^{(i)} \}$.
  \item superposition of the independently boosted boson stars
        \begin{eqnarray}\label{ID_equalcase}
           \phi &=& {\hat \phi}^{(1)}(\vec{x}-\vec{x_1}) + {\hat \phi}^{(2)}(\vec{x}-\vec{x_2}) \\
           g_{ab} &=& {\hat g}_{ab}^{(1)}(\vec{x}-\vec{x_1}) + {\hat g}_{ab}^{(2)}(\vec{x}-\vec{x_2})
	           - g_{ab}^{\rm Mink}
         \end{eqnarray}
\end{enumerate}
where $g_{ab}^{\rm Mink} = {\rm diag}(-1,1,1,1)$ is the flat metric in Cartesian coordinates.
The field $\phi^{(i)}$ (the unboosted scalar field) is dictated by the type of
boson star considered, that is, either a boson star or an anti-boson star. The prototype of the single boson star $(i)$ used throughout this work has $M^{(i)}=0.5$ and radius $R^{(i)}_{95}=24 M^{(i)}$ (defined as the radius
containing $95\%$ of the mass) , so its compactness  is comparable to a soft neutron star ($M^{(i)}/R^{(i)} \simeq 0.04$). We also note that these boson stars are field configurations of a single 
complex scalar field.


\section{Analysis quantities}
\label{analysis}

Throughout this work we define the center of each star as the location of a local maximum
of the energy density $\rho \equiv n^a\,n^b\,T_{ab}$ for each temporal slice. 
Due to the U(1) symmetry of the Lagrangian density (\ref{Lagrangian}),
there is a conserved Noether current defined by
\begin{equation}\label{noether_current}
  j^a = -\frac{i}{2} g^{ab} \left[\bar \phi ~\partial_{b} \phi-
  \phi~\partial_{b} \bar \phi \right].
\end{equation}
The conserved Noether charge $N$, associated with the number of bosonic
particles, can be expressed as
\begin{equation}\label{noether_charge}
     N = \int j^0~ \sqrt{-g}~ dx^3 ~~.
\end{equation}

The ADM mass and the angular momentum of the system are computed as
\begin{eqnarray}\label{defmass}
  M &=& \frac{1}{16\pi} \lim_{r \rightarrow \infty}
  \int g^{ij}~\left[\partial_j g_{ik} - \partial_k g_{ij}\right]~{\cal N}^k~dS \\
  J_i &=& \frac{1}{8\pi} \lim_{r \rightarrow \infty} {\epsilon_{il}}^m
  \int x^{l}~\left[K_{jm} - g_{jm}~{\rm tr}K \right]~{\cal N}^j~dS  
\end{eqnarray}
where ${\cal N}^k$ stands here for the unit outward normal to the sphere.

The gravitational radiation is described asymptotically by the Newman-Penrose $\Psi_4$ scalar.
To analyze the structure of the radiated waveforms it is convenient
to  decompose the signal into $-2$ spin weighted spherical harmonics as
\begin{equation}\label{rpsi4}
  M~r~\Psi_4 = \sum_{l,m} C_{l,m} {}^{-2} Y_{l,m}
\end{equation}
where the factor $r$ is included to better capture the $1/r$ leading order behavior of $\Psi_4$.

We also focus on integral quantities that are independent of the specific
basis of the spherical harmonics, such as the radiated energy and the radiated angular
momentum. Using both the decomposition (\ref{rpsi4}) and the orthonormalization of the spherical harmonics, these expressions can be written as
\begin{eqnarray}\label{dEdt}
  \frac{dE}{dt} &=& \frac{1}{16\pi} \sum_{l,m} | D_{l,m}(t) |^2 ~~,\\
  \frac{d J_z}{dt} &=& - \frac{M}{16\pi} \sum_{l,m} m~({\rm Im}[D_{l,m}(t)~E^{*}_{l,m}(t)]) ~,
\end{eqnarray}
where we have used the adimensional quantities
\begin{eqnarray}\label{dEdt2}
  D_{l,m} &=& \frac{1}{M} \int^t_{-\infty} C_{l,m}(t')~ dt' \\
  E_{l,m} &=& \frac{1}{M} \int^t_{-\infty} D_{l,m}(t')~ dt'.
\end{eqnarray}


\section{Simulations and Results}
\label{numerics}
Our starting point is an equal-mass binary system with initial
velocities and separation chosen so that a simple Newtonian analysis would
predict the system to be bound.
While both stars have equal mass, we exploit the freedom in setting
the sign of the frequency and the phase of the stars to concentrate primarily on two
different cases. 
The first case consists of two identical boson stars (the BB case), and the other consists
of a boson star paired with its antiboson partner (the BaB case).
One could also consider a boson star interacting with a copy of itself offset in phase, but we defer
this to future work. Here we begin by surveying the dynamical behavior of the BB case
as its angular momentum is varied. This allows us to examine the possible phenomenology
of the system. Next we concentrate on two particular cases with total initial orbital angular 
momentum (oriented along the $z$ axis) given either by $J_z = 0.8 M^{2}$ or $J_z = 1.1 M^{2}$. 
These values have been chosen so that the angular momentum is either below or just above
the lowest allowed quantized value of the (rotating) stationary axisymmetric boson star, 
which obeys
\begin{equation}\label{quantized}
   J_z = n~N
\end{equation}
with $n$ an integer. Roughly speaking,  if $N^{(i)}$ denotes the Noether charge
for a stable star (where $i=1$ denotes the first star and $i=2$ the other), then its mass is approximately given by
 $M^{(i)} \simeq \epsilon_i~N^{(i)}/m$, being $\epsilon_i=\pm 1$ whether it is a boson or an antiboson star
 and $m$ the boson mass (see for instance Fig. 3 of~\cite{YosEri97}). 
As a result, the Noether charge evaluated on the initial hypersurface is approximately 
\begin{equation}
  N = \frac{M}{m} = \frac{1}{m} \left( \epsilon_1 M^{(1)} + \epsilon_2 M^{(2)} \right) .
\end{equation}
Here we consider $m=1$ and $M^{(1)}=M^{(2)}=0.5$, so that the initial Noether charge
for the boson/boson case is $N \simeq M^2$  while for the boson/antiboson pair we have precisely $N = 0$,
which can be exploited to analyze the possible outcomes of a merger (or close interaction) of the stars.

In the BaB case, since the total Noether charge must be zero, the interaction can not give
rise to a single stationary regular object, so the final stage can not be the spinning boson star described
in~\cite{YosEri97}. A more intuitive explanation relies on observing that a prospective
remnant boson star would no more likely have positive  charge than negative and so cannot exist.
Therefore, either annihilation, dispersion or the (unlikely) production of multiple pairs
of boson/antiboson stars is to be expected in the cases where a black hole is not formed.

In the BB case, the Noether charge is non-zero, which by itself does not rule out the
possibility of a single regular object as the final outcome.
However, angular momentum considerations do shed further light on this issue. Notice that 
the boson/boson binary cases we consider have orbital angular momentum with values slightly 
above and below the Noether charge.  In the latter case, since the total angular
momentum is below the minimum allowed value, an outcome describing a single star with a mass comparable to the total mass would seem impossible. In the former case, while forming a single boson star is possible,
the system must shed enough angular momentum/energy to satisfy the relation (\ref{quantized}). 
As we will see below, the dynamics of the system reveals a
a rather involved behavior.  We arrange our presentation so as to discuss
the case with smaller angular momentum first, and then concentrate on the higher momentum
case. 

We study the dynamics of these two cases using finite differences to
approximate the equations of motion within a distributed and adaptive infrastructure.
For the simulations of the angular momentum survey in Section~\ref{sec:overview},
the computational domain covers the region $x^i\in [-200M,200M]$ while the stars begin within about
$40M$ of each other, where $M \simeq 1$ is the total ADM mass. This separation, which is larger
than what we adopt in the following subsections,
allows for a larger angular momentum in a bounded binary system and thus our survey is more exhaustive.
Within this domain, finer grids are placed dynamically according to an estimate of truncation
error provided by a self-shadow hierarchy. Typically, refined regions contain both stars and
the span between them with even finer regions tracking the individual stars providing
a minimum grid spacing for each star of $\Delta x=0.125M$ and a maximum grid spacing far from
the stars of $\Delta x=4.0M$.

For the other simulations, the computational domain extends to $x^i\in [-400M,400M]$ although the stars begin within about $32M$ of each other. Such a large coarse grid allows us to extract
gravitational radiation far from the dynamics. We compute surface integrals (e.g. the ADM mass) and $\Psi_4$ at extraction surfaces located at $r_{\rm ext}= 150M,~180M$ and $210M$,
where the grid spacing is given always by $\Delta x=2M$.


\subsection{Survey results}
\label{sec:overview}

In this subsection we describe briefly the dynamics found
in the binary boson star system as we vary the initial velocities of the stars.
Here, we make a few general remarks about these results,
but we should note that we have yet to fully explore the
parameter space. However in the limited region so far explored,
a rich phenomenology is uncovered by our simulations.
We choose stars with masses $M^{(i)}=0.5$
near the maximum mass allowed on the stable branch $M_{\rm max}=0.63$. Further,
while the stars begin centered at $(x,y,z)=(0, \pm 16M, 0)$, their initial speeds
are perpendicular to their relative position vector,
$(v_x,v_y,v_z)=(\pm v_{\rm boost}, 0, 0)$.

We find that for low speeds,
the stars approach and form a spinning black hole as one
would expect for two bound massive compact objects which, in isolation,
are already close to the unstable branch.
For stars with very large initial speeds, the system is
unbound and the stars continue to fly away from each other.

The interest lies in the intermediate range of speeds for
which the stars interact nonlinearly. In this regime, as one
increases the initial velocities the stars merge into a short-lived
rotating object which finally
splits into two pieces that are ejected.
Continuing to increase the initial speeds, one finds the ejecta
decreasing in size with a concomitant increase in dispersion of scalar field.
Eventually as one increases the velocities one observes the formation of
a central black hole.

\begin{figure}[h]
\begin{center}
\epsfxsize=8cm
\epsfbox{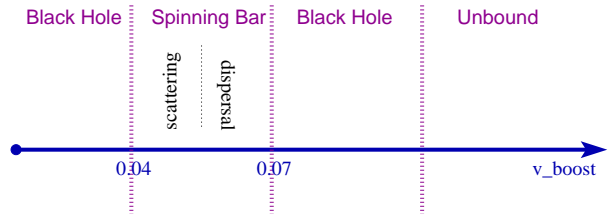}
\end{center}
\caption{Phenomenology for the boson/boson interaction with angular momentum (specific to the family of initial data described in the text). There are
two main different behaviours in the bound case, the formation of either a black hole or 
a rotating bar. The rotating bar, depending on the angular momentum, will either split into
two objects or it will disperse on a longer time scale. Animations
obtained from the simulations 
for some of these cases are available online~\cite{online}.}
\label{overview_bb}
\end{figure}

The various transitions in parameter space just described are sketched
in Fig.~\ref{overview_bb} (and representative animations 
can be found online~\cite{online}). Such phenomenology demands
more thorough study. However,
we defer to a future work this more detailed investigation as the
associated computational costs are very high. In the present work
our focus is to obtain the gravitational
wave signature for representative cases and compare the results with
those of other compact orbiting systems.

\subsection{Small Angular Momentum ($J_z = 0.8 M^{2}$)}

\subsection*{Boson/Boson pair}

We consider a binary boson/boson star system with initial angular momentum of $J_z=0.8~M^2$.
As the evolution proceeds, the
stars approach, orbit about each other for about half an orbit before merging into a 
single object. This object however, after spinning as a single entity for about $t\simeq 200M$,
splits into two identical objects which fly apart from each other. Thus, this case
corresponds to what we have labeled as the spinning bar/scattering region of Fig. ~\ref{overview_bb}.
 
\begin{figure}[h]
\begin{center}
\epsfxsize=7cm
\epsfbox{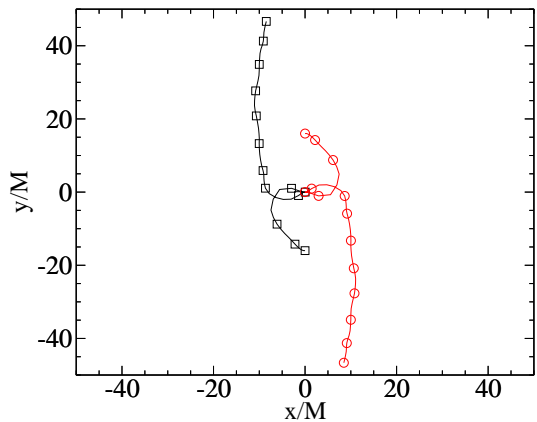}
\end{center}
\caption{\textit{Boson/boson pair ($J_z=0.8 M^2$)}. The position of the center of the star, centered
initially at $(x,y)=(0, \pm 16M)$,  during the merge and the scattering. The stars form a single rotating
object for $T=200M$ before splitting in two equal objects that are ejected towards infinity.}
\label{center_unmerge}
\end{figure}

The trajectories are plotted in Fig.~\ref{center_unmerge}, where
the coordinate position of the stars' centers are shown for different times.
A careful inspection of the scattered stars indicate their mass is half what 
the initial stars have. Furthermore, a Newtonian
calculation would indicate that the final speeds of the stars are above the escape speed if
no other gravitating source were present. 
Most of the remaining mass was dispersed during the merger and spinning, and, although a small fraction
crosses the extraction surfaces, most of the scalar field is found in an extended halo surrounding the
central merger region. 

\begin{figure}
\centering
\begin{tabular}{c}
\epsfig{file=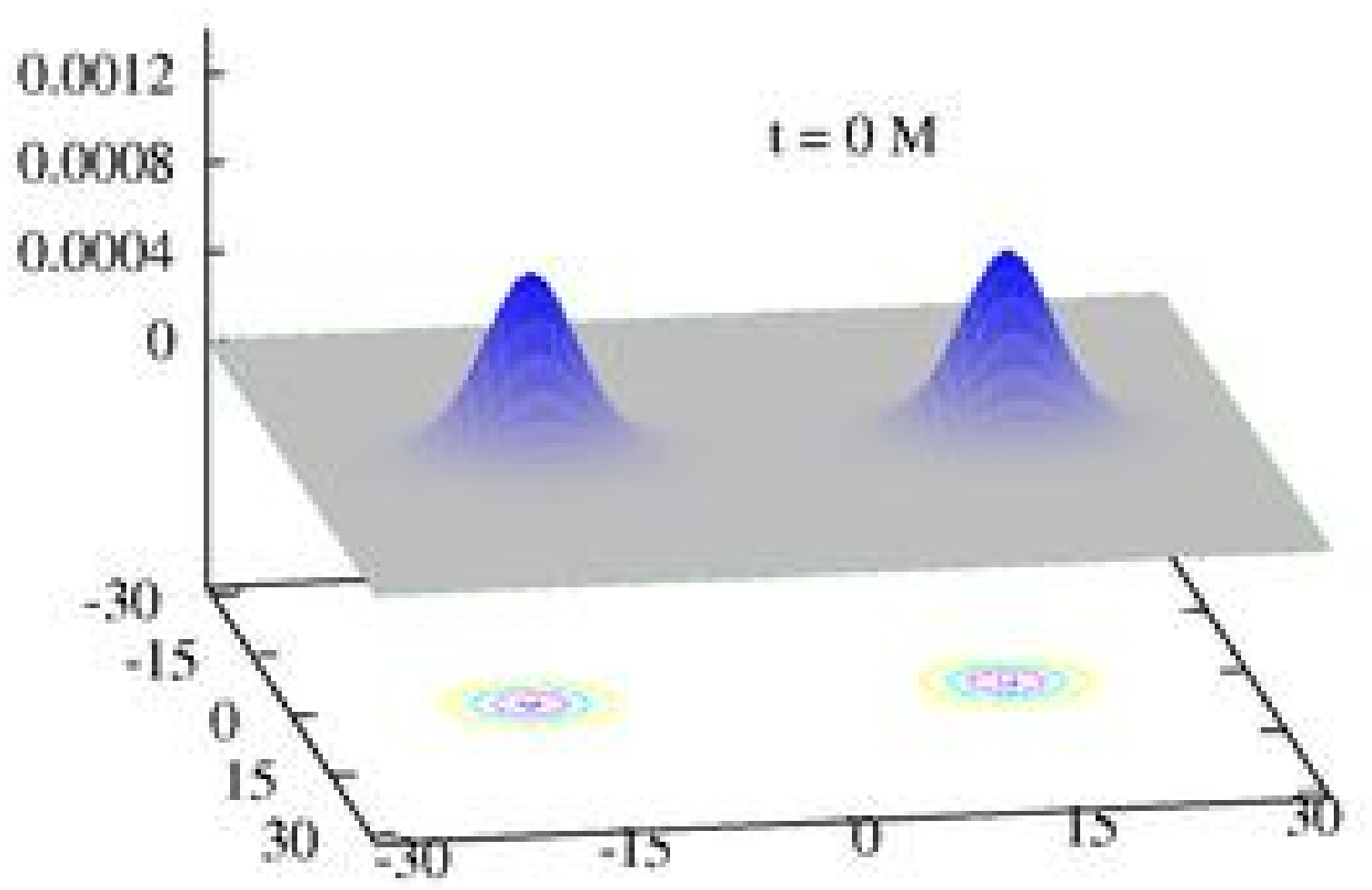,width=0.75\linewidth,clip=} \\
\epsfig{file=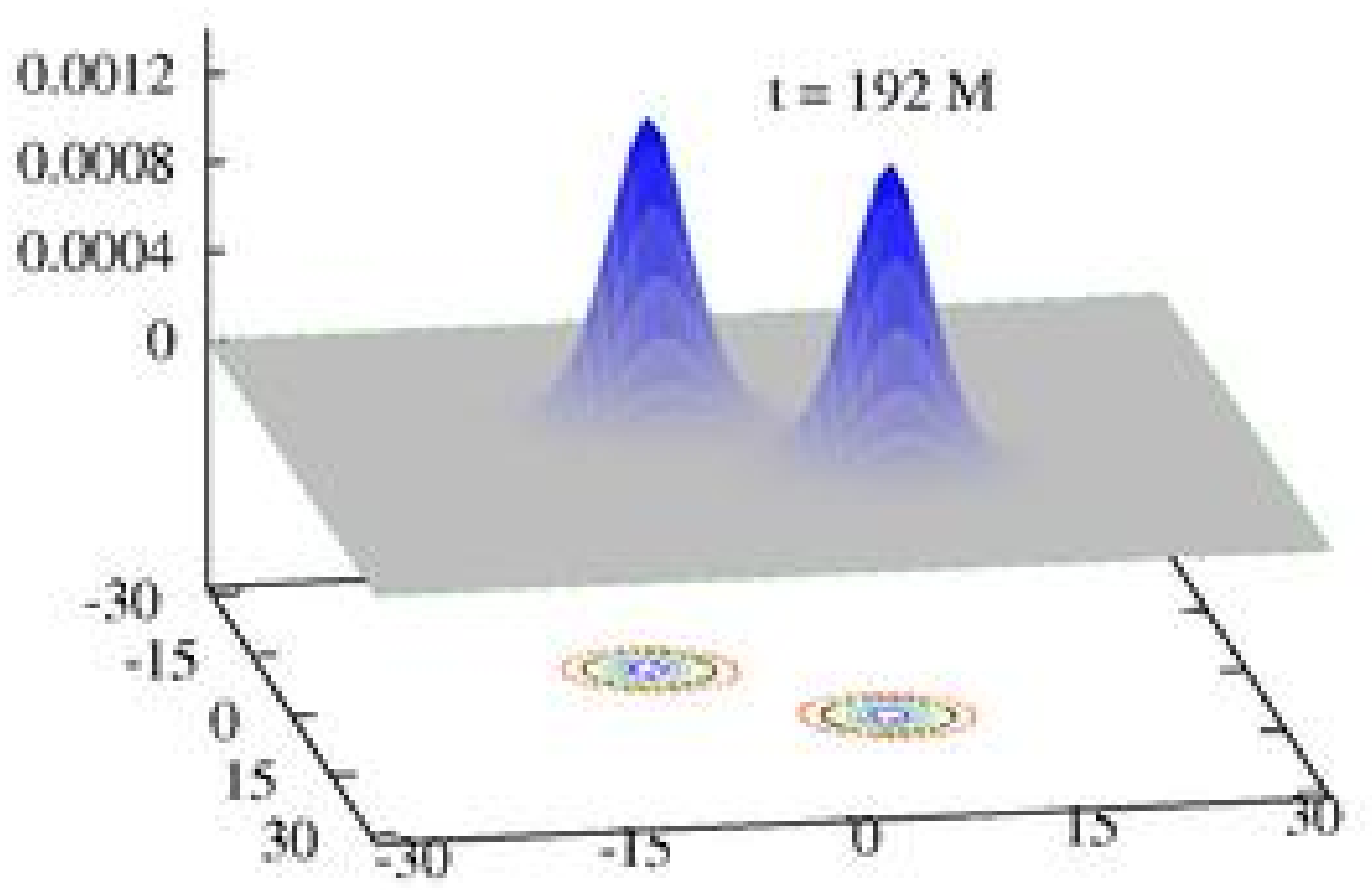,width=0.75\linewidth,clip=} \\
\epsfig{file=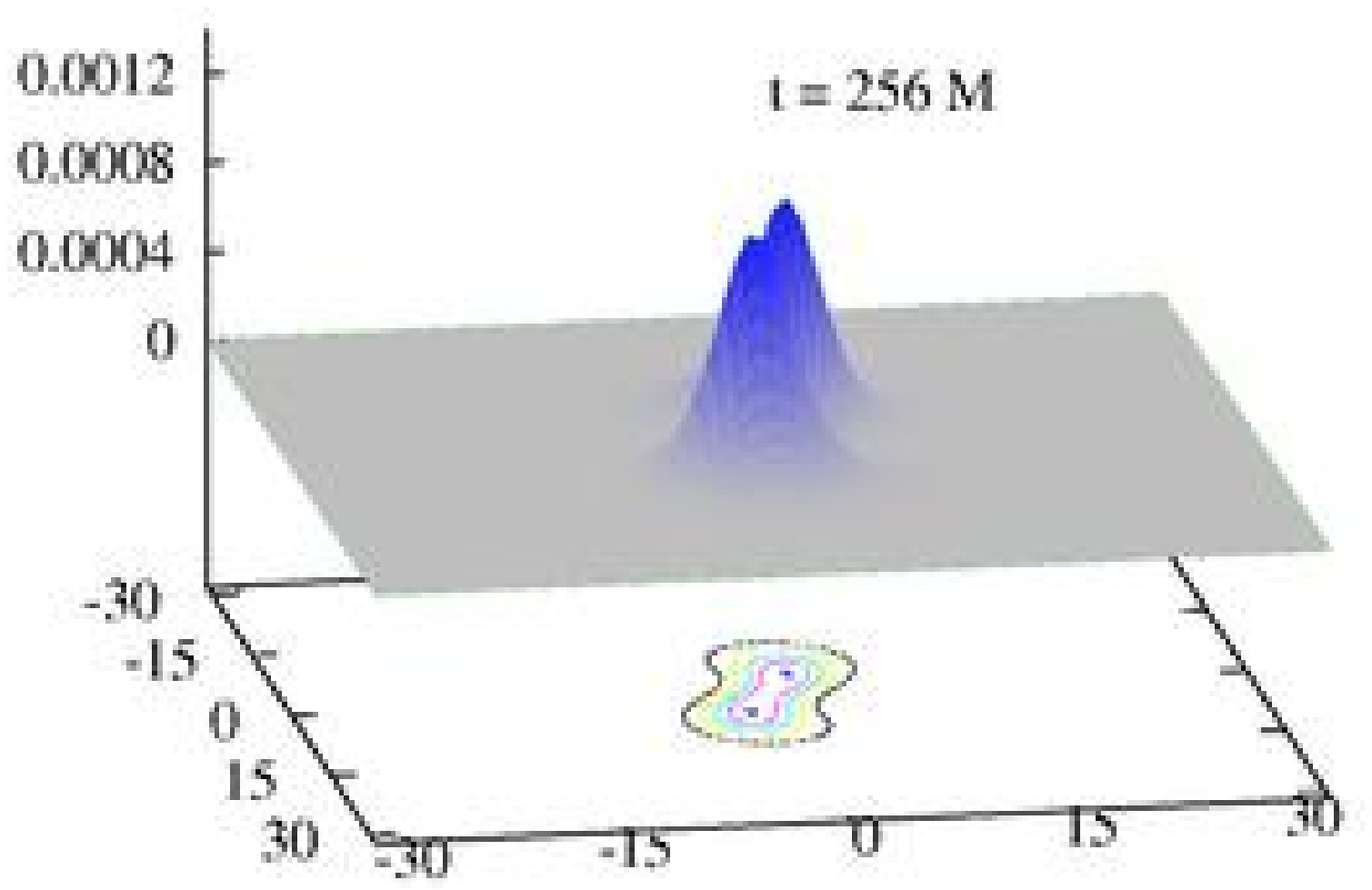,width=0.75\linewidth,clip=} \\
\epsfig{file=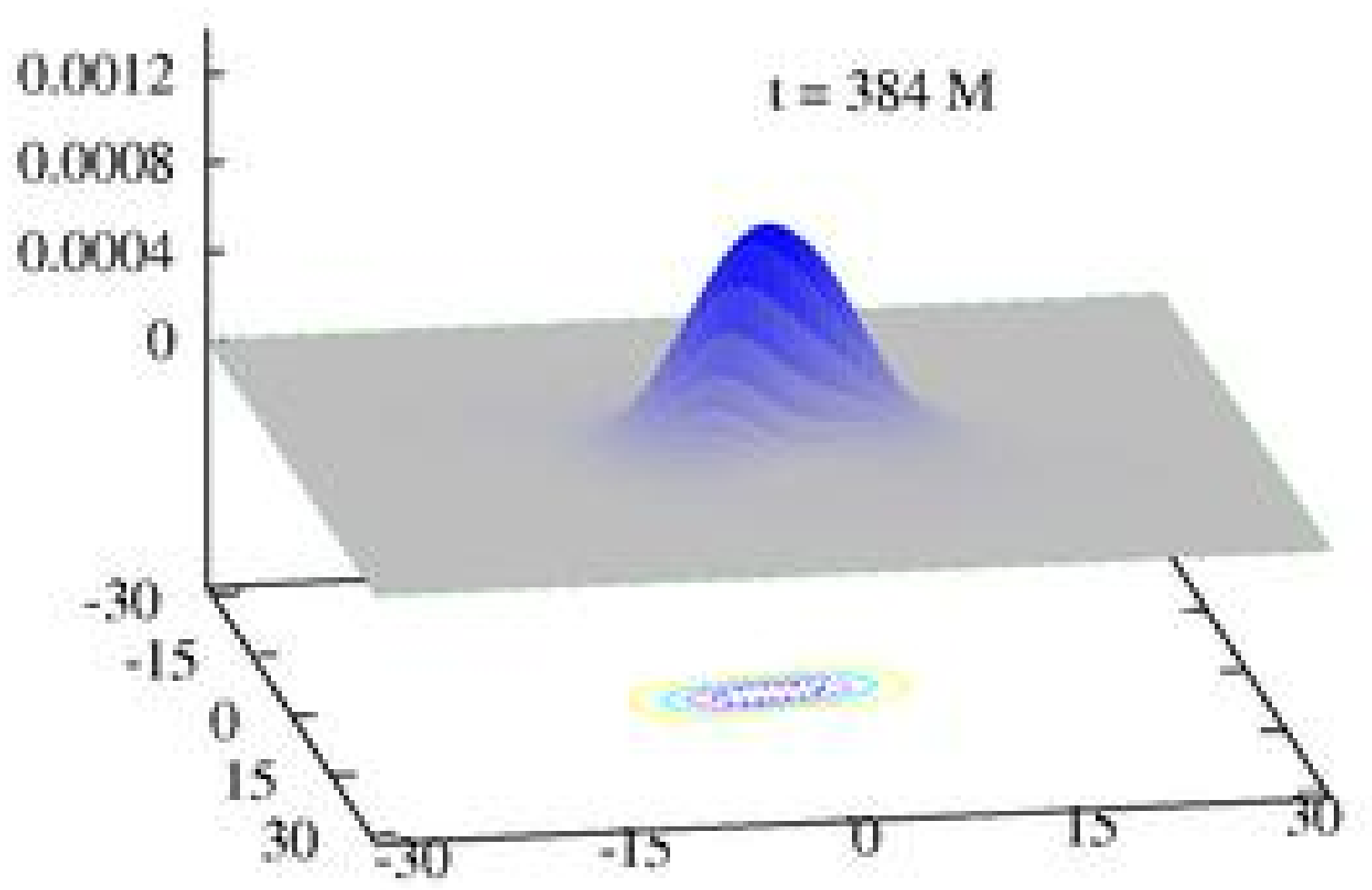,width=0.75\linewidth,clip=} \\
\epsfig{file=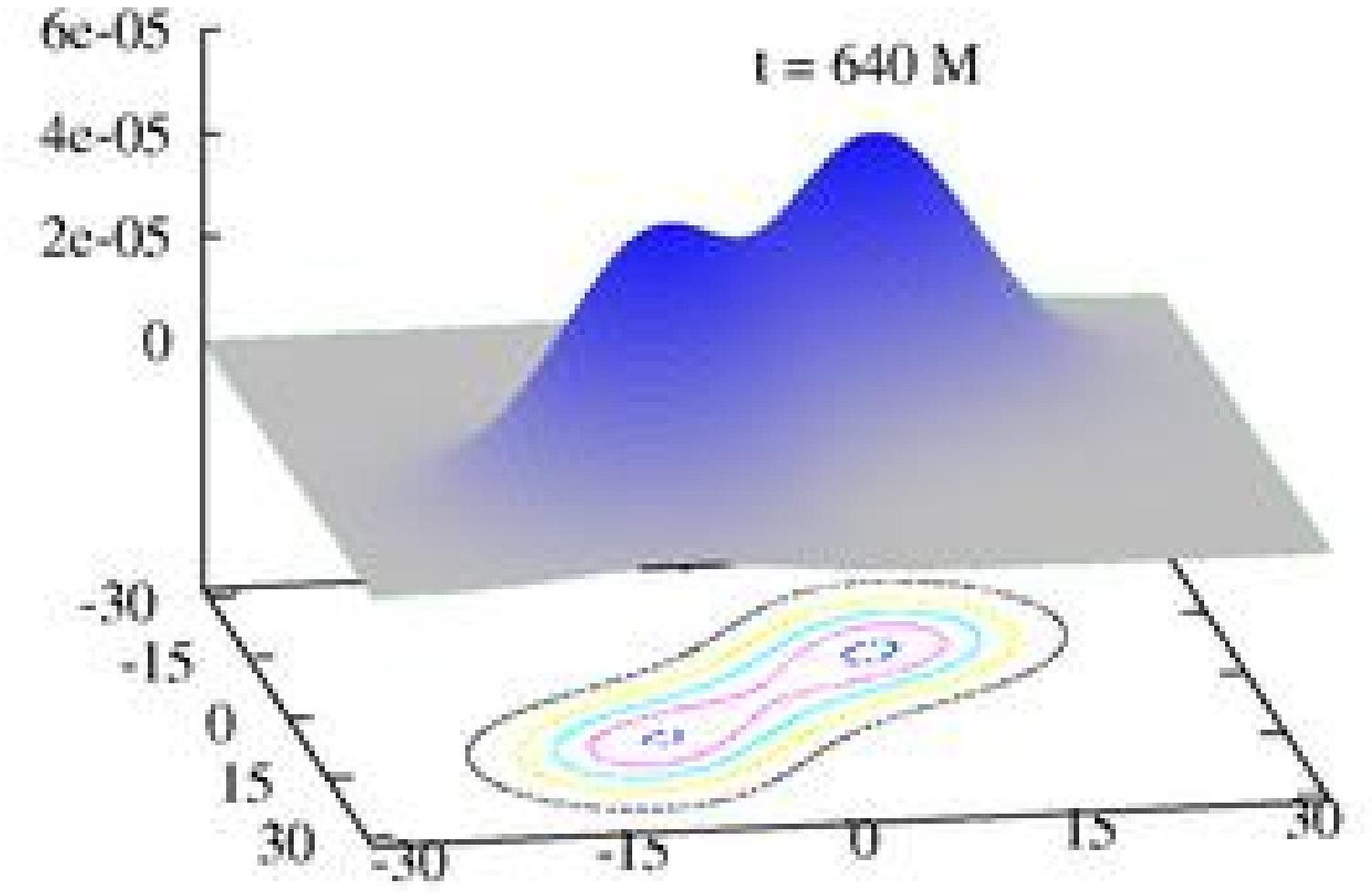,width=0.75\linewidth,clip=}
\end{tabular}
\caption{\textit{Boson/boson pair ($J_z=0.8 M^2$)}. Snapshots at the $z=0$ plane of the energy
density $\rho$ for different times. As the stars come closer and merge, the maximum value of $\rho$ grows significantly. The boson stars keep together for $t=200M$ and then they fly away with significantly less mass than initially.}\label{tau_unmerge}
\end{figure}

Therefore, the system can be regarded as undergoing a non-trivial scattering
in which the initial stars give rise to an unstable merged object.
The instability of this object likely results from having
insufficient angular momentum to settle into even the lowest allowed rotating, stationary boson star.
As a result the smaller stars are `kicked-out' in a configuration whose angular momentum is slightly below
the initial one. 
The details of this scattering can be seen in Fig.~\ref{tau_unmerge}, where several snapshots of the
energy density at different times are plotted. 
The remnant scalar field remains in an extended halo around the origin.

\begin{figure}[h]
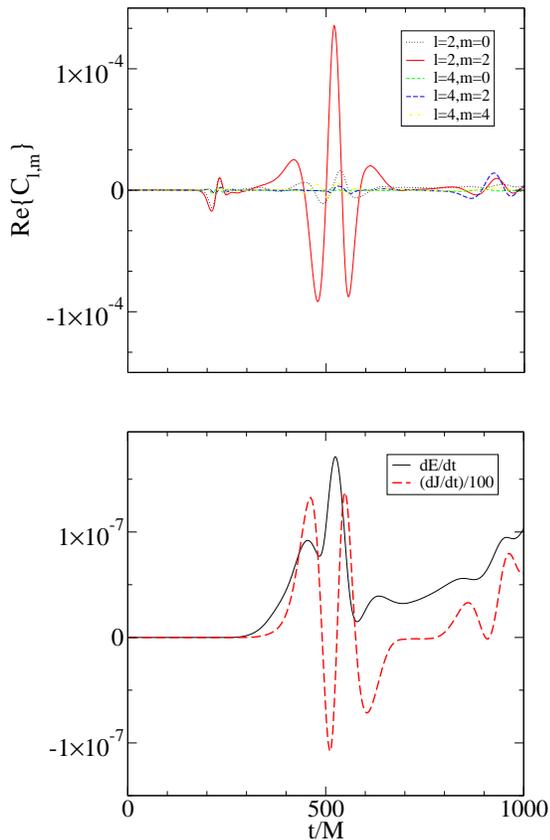

\begin{center}
\epsfxsize=2.6in
\epsffile[40 18 707 584]{psi4_all_unmerge.eps}\\
\epsfxsize=2.6in
\epsffile[40 59 708 584]{EJ_unmerge.eps}\\
\end{center}
\caption{\textit{Boson/boson pair ($J_z=0.8 M^2$)}. Different modes of the $\Psi_4$ (normalized
by $M~r$) at the extraction radius $r=210M$ as a function of time at the top and
the flux of energy and angular momentum at the bottom. Notice that the flux of angular momentum
is about two orders of magnitude larger than the flux of energy.}
\label{psi4_all_unmerge}
\end{figure}

The emitted waveforms show a dominant $l=2,m=2$ mode in the spin-weighted decomposition of $M r \Psi_4$
computed on the surface extraction at $r=210M$, as demonstrated in Fig.~\ref{psi4_all_unmerge}. 
Its qualitative features are reminiscent of the waveforms produced in zoom-whirl orbits of
compact objects\cite{hughes}. The flux of energy and angular momentum are also plotted in Fig.~\ref{psi4_all_unmerge}.

Because the stars are rapidly moving towards the extraction surfaces, and errors are propagating
from the outer boundary, the waveforms become untrustworthy
after $t=800$. Nevertheless the dominant portion of the gravitational wave output
in the system is clearly seen. Additionally, as indicated in the figure, 
the amount of angular momentum radiated is significantly stronger than that of
the radiated energy, although both of them represent a small fraction of the initial quantities.

\begin{figure}[h]
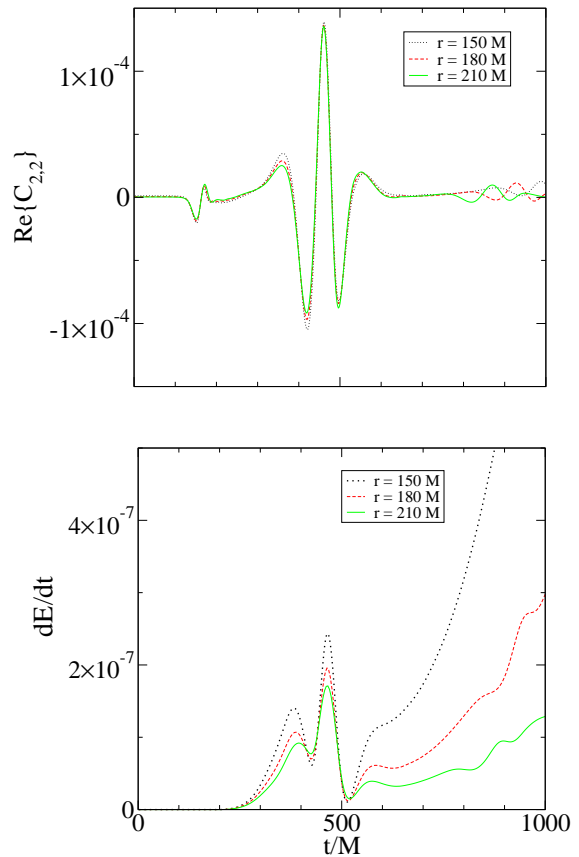

\begin{center}
\epsfxsize=2.7in
\epsffile[40 18 707 584]{psi4_r_unmerge.eps}\\
\epsfxsize=2.7in
\epsffile[10 29 708 584]{fluxE_r.eps}\\
\end{center}
\caption{\textit{Boson/boson pair ($J_z=0.8 M^2$)}. The dominant mode $C_{2,2}$ of $M r \Psi_4$ 
and the flux of energy $dE/dt$ extracted at spheres with different radii. 
The plots have been shifted horizontally by the radial distance between the spheres.}
\label{psi4_r_unmerge}
\end{figure}

Additionally, this dominant mode and the energy flux are computed at different
extraction radii in Fig.~\ref{psi4_r_unmerge} (top), showing the expected wave-like behaviour of
the Weyl scalar $\Psi_4$ and a convergent behaviour of the energy flux as the extraction radius
is more distant. The effects of having the surface extraction too close to the stars
are shown in the same figure (bottom), where the energy flux is plotted for the three
different radius. The tail of this quantity, much more sensitive than $\Psi_4$ to distance
effects, converges to zero as the extraction surface is placed farther away.

\begin{figure}[h]
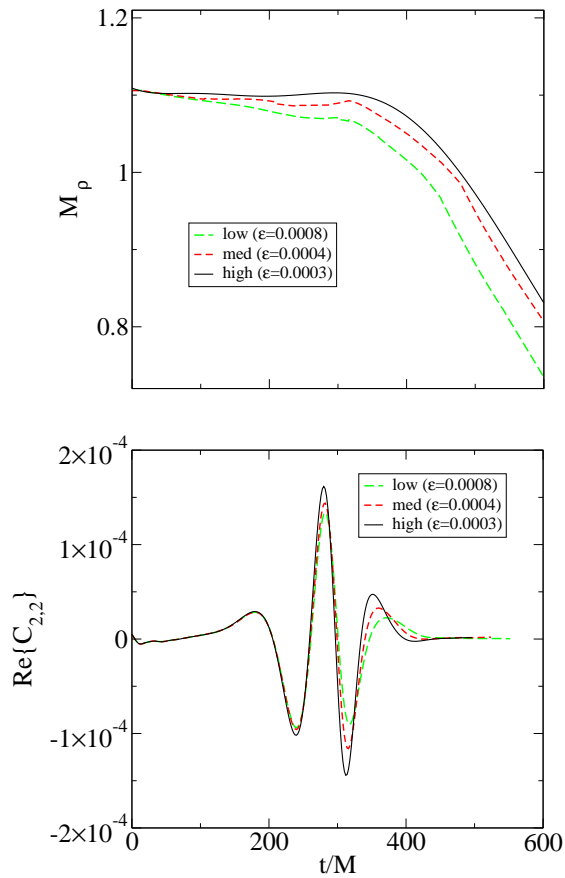

\begin{center}
\epsfxsize=2.7in
\epsffile[40 18 707 584]{convergence_mass.eps}\\
\epsfxsize=2.7in
\epsffile[40 29 708 584]{convergence_c2p2.eps}\\
\end{center}
\caption{\textit{Boson/boson pair ($J_z=0.8 M^2$)}. The integral of the energy density
$M_{\rho}$ and the dominant mode $C_{2,2}$ of $M r \Psi_4$ (extracted at $r=210M$ and
shifted horizontally by $t=210M$) for three different truncation error thresholds $\epsilon$.}
\label{convergence}
\end{figure}

In our previous work, we presented convergence tests in which the resolution
was increased by a fixed amount. These tests indicated the code converged as
expected. Here, as a complementary test, we examine the solution's behavior when we vary the
refinement criterion. In particular, when the estimate of the truncation error
exceeds a user defined threshold $\epsilon$, refinement is added there.
Notice that this test is not designed
to examine the convergence rate of the solution, but rather to study the 
implementation's ability 
to adjust the grid structure so that the solution's error
stays below a given one. By decreasing the value of $\epsilon$ the grid
structure effectively increases the resolution where needed. 
Fig. ~\ref{convergence} illustrates the integral of the energy density and the
principal mode of the $\Psi_4$ decomposition for three different values of $\epsilon$. 
As is evident from these graphs, the quality of the obtained solution improves as
the threshold is decreased and displays convergent behavior.  

\subsection*{Boson/Antiboson Pair}

In contrast to the BB case, the total Noether charge for the boson star and its antiboson
partner is zero. One implication of this cancellation is that the system cannot settle into
a single remnant star. In spite of this preclusion, we observe the stars merging into some sort of
asymmetric, inhomogeneous rotating object.

\begin{figure}
\centering
\begin{tabular}{c}
\epsfig{file=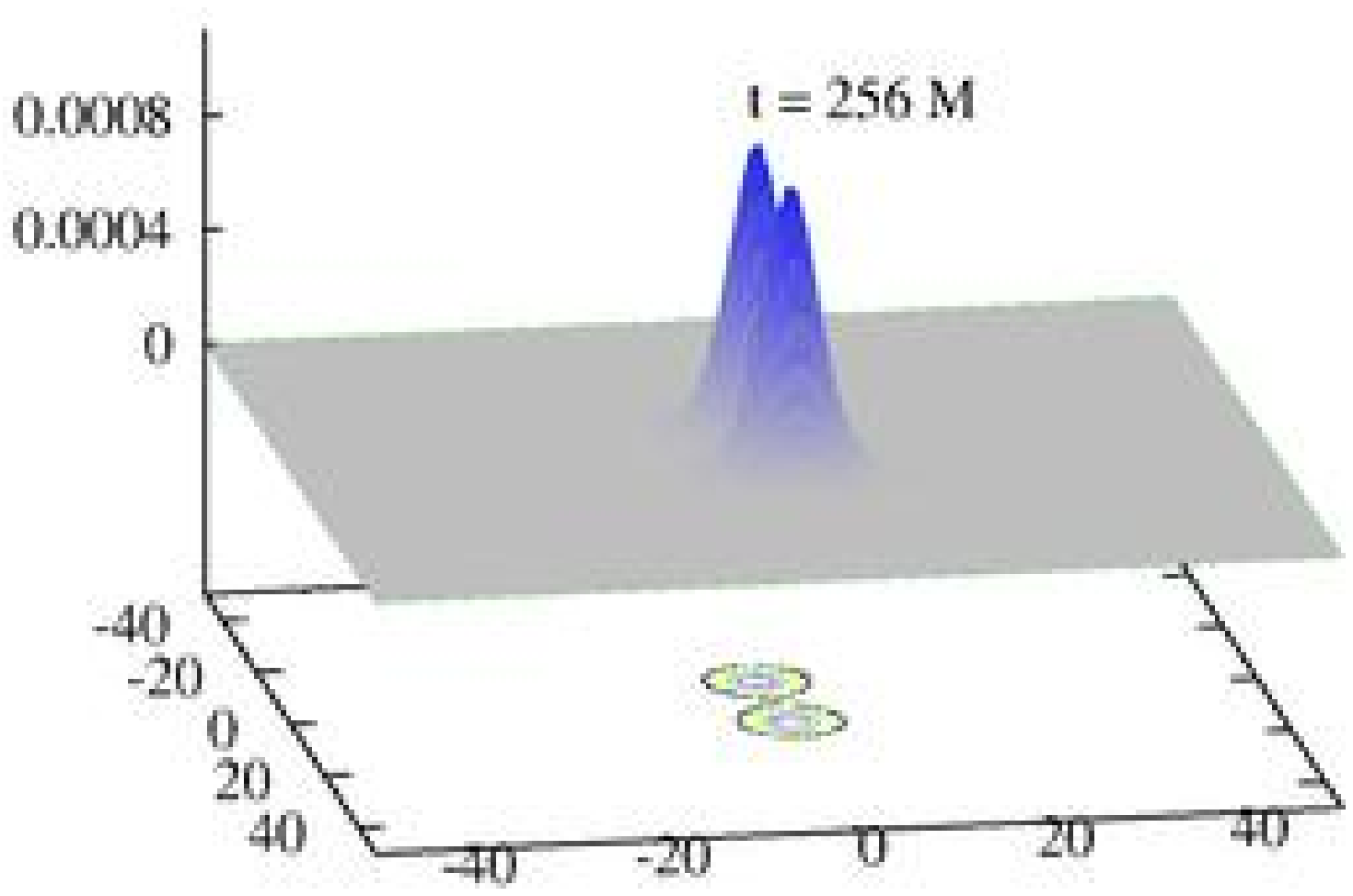,width=0.75\linewidth,clip=} \\
\epsfig{file=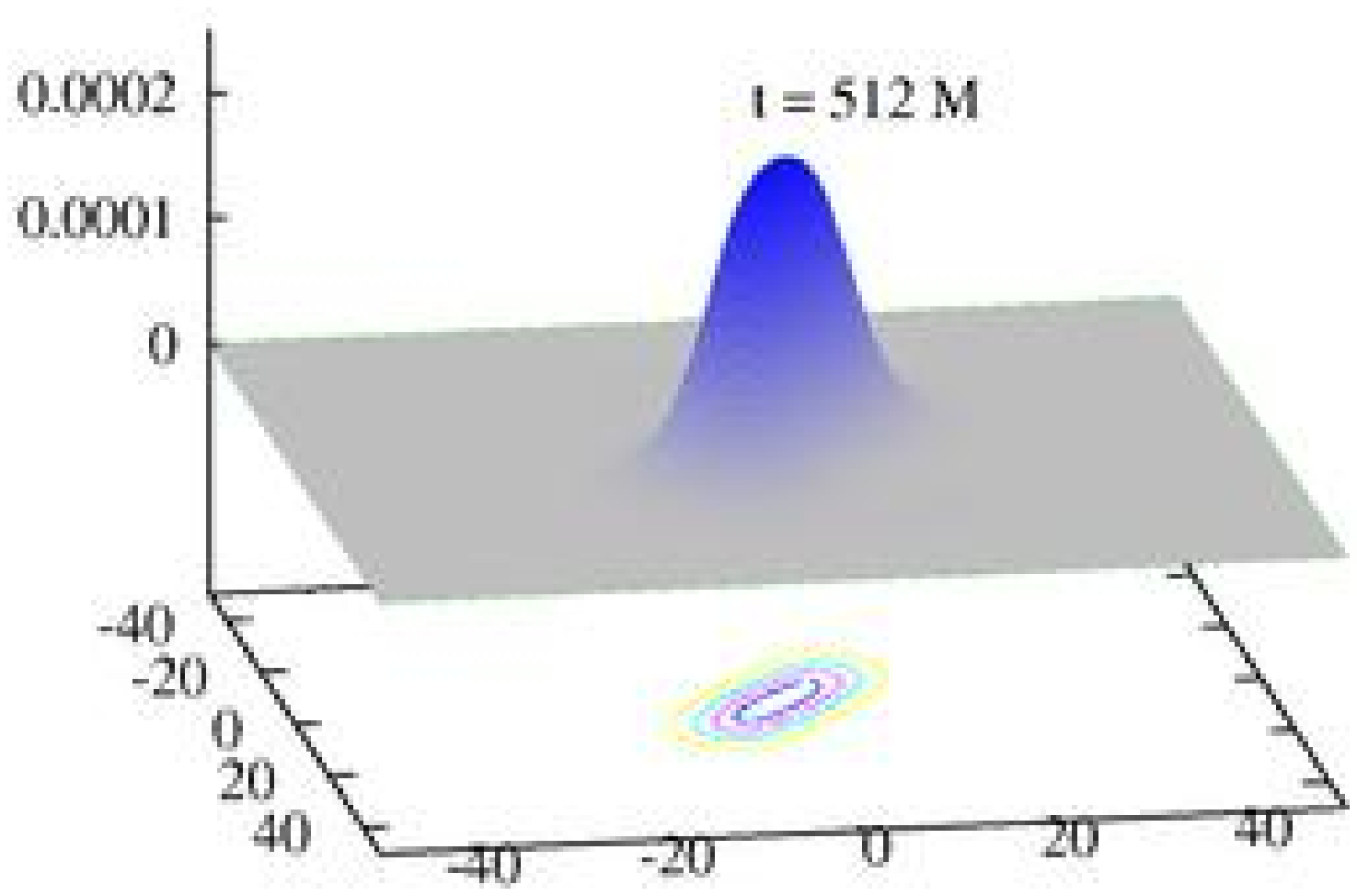,width=0.75\linewidth,clip=} \\
\epsfig{file=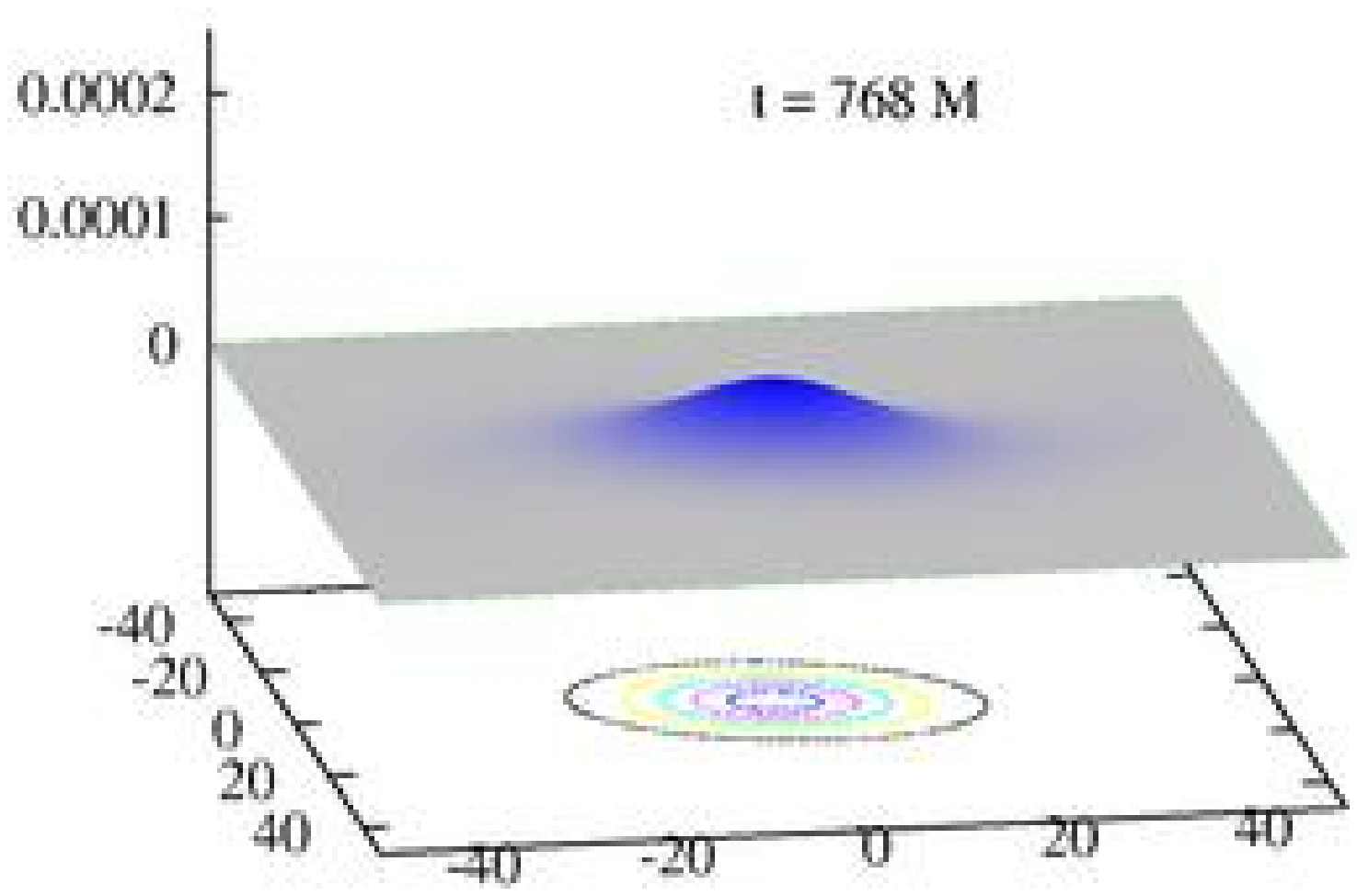,width=0.75\linewidth,clip=} \\
\epsfig{file=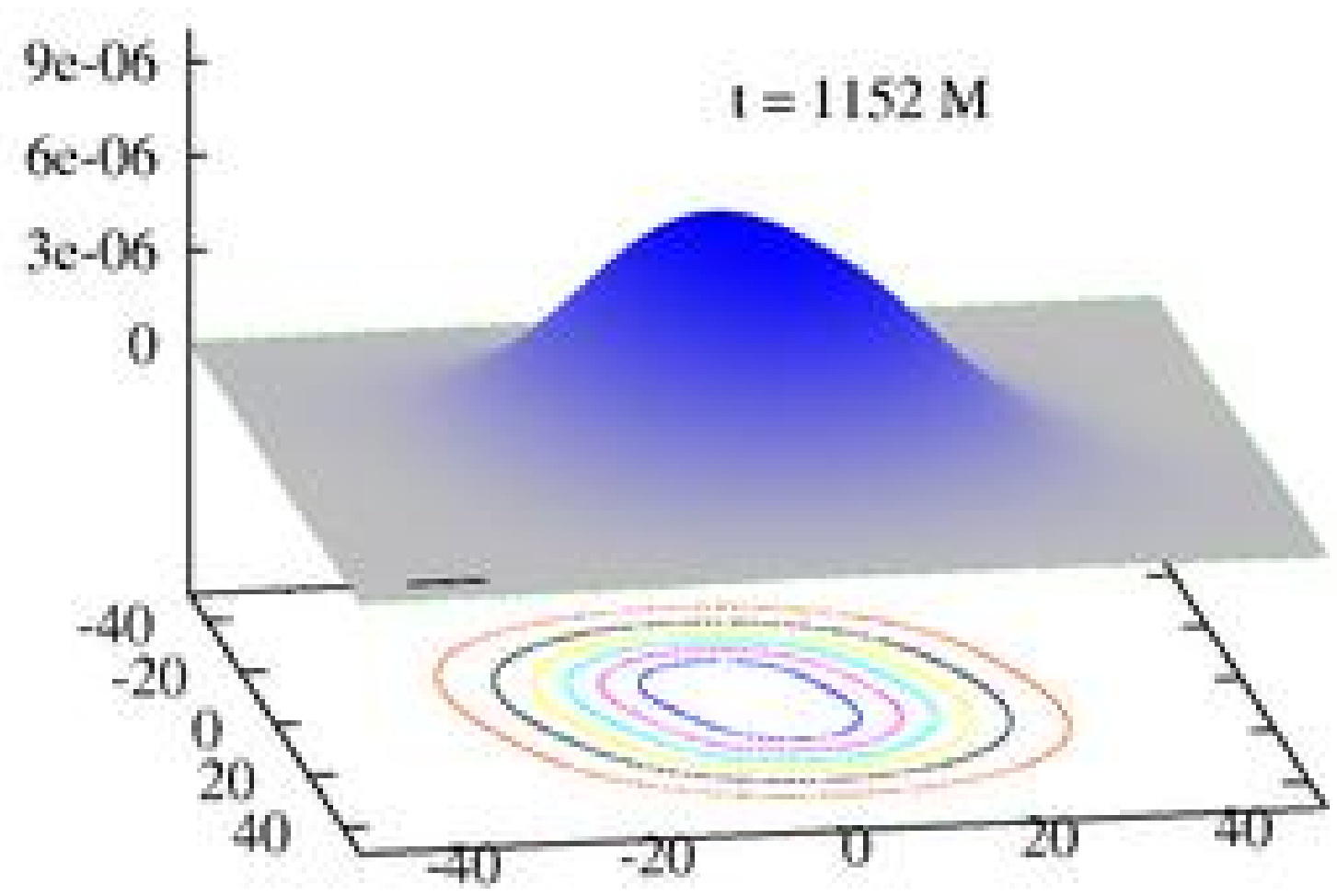,width=0.75\linewidth,clip=}
\end{tabular}
\caption{\textit{Boson/antiboson pair ($J_z=0.8 M^2$)}. Snapshots at the $z=0$ plane of
the energy density $\rho$ for different times. Notice that the maximum of $\rho$ decreases while
the radius of the star increase.}
\label{tau_anhi}
\end{figure}

\begin{figure}[h]
\begin{center}
\epsfxsize=8cm
\epsfbox{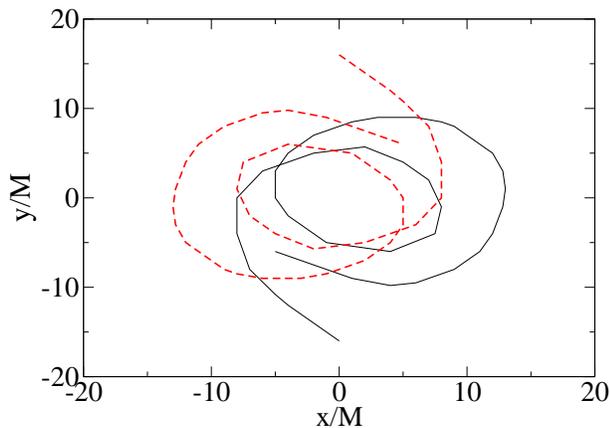}
\end{center}
\caption{\textit{Boson/antiboson pair ($J_z=0.8 M^2$)}. The position of the extremes of the Noether
density, centered initially at $(x,y)=(0, \pm 16M)$, which define roughly the location of
the boson and the antiboson.}
\label{center_anih}
\end{figure}

Starting from an initial configuration identical to the one from the
previous subsection  with the exception that one star is an antiboson star, the behavior is illustrated
in Fig.~\ref{tau_anhi}. Snapshots of the energy density show
the stars orbiting very close while much of the scalar field 
disperses from the origin. The system settles into a larger inhomogeneous object, 
apparently a rotating composite of a 
boson and antiboson star. The trajectories of these components are plotted in Fig.~\ref{center_anih}.

The radius of this object, computed from the energy density, varies with time.
For instance, at about $t\simeq 1000M$ its size has grown to about five times 
the size it had shortly after the merger, $L_0$. The expansion continues to about
$t\simeq 1500M$ when its size is about $6 L_0$. At this point the expansion halts
and turns around, decreasing its size to approximately $3 L_0$. Afterwards
this size stays roughly the same. Most of the scalar field energy density does not disperse away but 
remains contained in a big sphere centered at the origin. Although we can see that the
energy density crossing the surface extraction is about two orders of magnitude larger than in the
BB pair with the same angular momentum, the escaped scalar field is still a small portion
of the total one.

\begin{figure}[h]
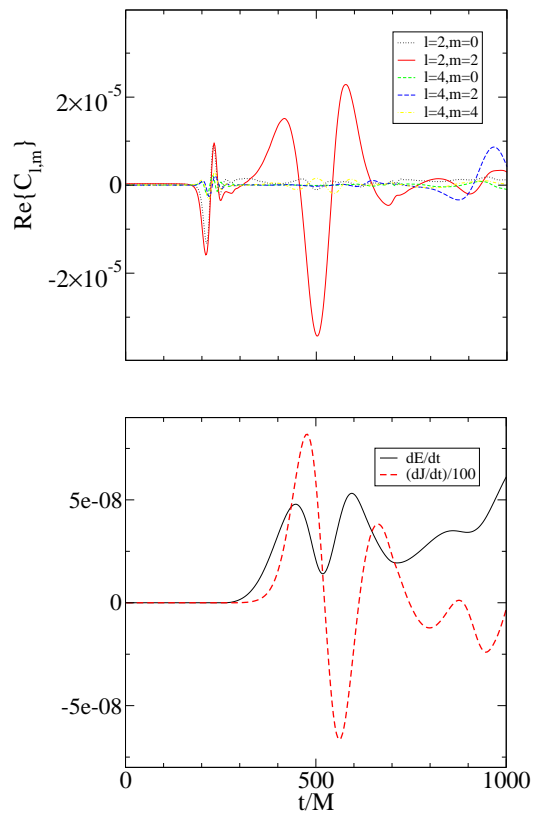

\begin{center}
\epsfxsize=2.5in
\epsffile[40 18 707 584]{psi4_anhi.eps}\\
\epsfxsize=2.5in
\epsffile[40 59 708 584]{EJ_anhi.eps}\\
\end{center}
\caption{\textit{Boson/antiboson pair ($J_z=0.8 M^2$)}. The principal modes of $M r \Psi_4$ 
at the further extraction radius ($r=210M$) as a function of time at the top and
the flux of energy and angular momentum at the bottom. The reflections and boundary effects become
dominant around $t=1000M$, prevent us to characterize the radiation of the final inhomogeneous
spinning object.}
\label{psi4_anhi}
\end{figure}

As before, we can compute $\Psi_4$ and decompose it into spin-weighted spherical 
harmonics. In Fig.~\ref{psi4_anhi} the lowest modes are plotted, extracted at
$r_{\rm ext}=150M$, showing again a clear $l=2,m=2$ dominant mode.
The energy and angular momentum fluxes are also plotted in the same figure, where it is clear
that the pattern is different from the BB pair in Fig.~\ref{psi4_all_unmerge}. Additionally,
once again, the flux of angular momentum is two orders of
magnitude larger than the energy flux.  

\subsection{Large Angular Momentum ($J_z = 1.1 M^{2}$)}

\label{orbitingcase}

Here we consider both the case of two boson stars orbiting each other (the BB case) and
a boson star and its antiboson partner (the BAB case) for large initial angular
momentum. In contrast with the previous case, here the BB pair has angular momentum
which exceeds the Noether charge, thus a possible end state is a single boson star with
angular momentum. Alternatively, a spinning black hole could be produced. Since
$J_z \simeq M^2$ this is a delicate outcome because this would be close to an extremal
Kerr black hole, unless a significant amount of angular momentum is shed.

The situation with the BaB case, while less involved, still has interesting
features. Since the Noether charge in this case is zero, a single boson 
star cannot be produced. Thus one expects that  either a black hole forms or the scalar field
disperses in such a way that the total angular momentum, up to the typically
small radiated amount, is accounted for. As we will see, the simulations reveal the resulting outcome
is the formation of a black hole.

\begin{figure}[h]
\begin{center}
\epsfxsize=8cm
\epsfbox{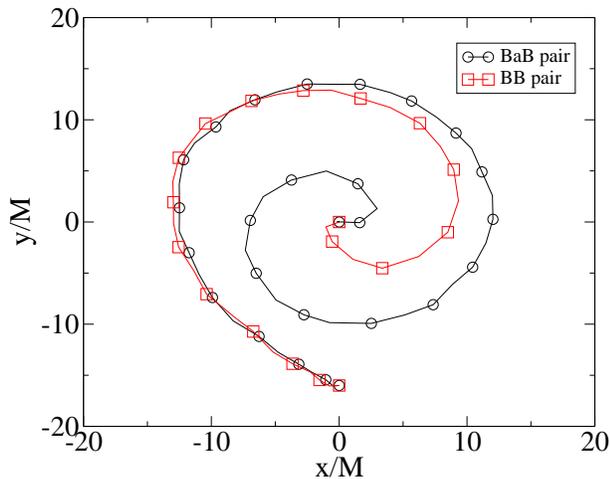}
\end{center}
\caption{\textit{Boson/boson and Boson/antiboson pairs ($J_z=1.1 M^2$)}. The coordinate position
of the Galilean-boosted boson stars for the BB and the BAB pairs.
The stars complete roughly one orbit before merging.}
\label{centers}
\end{figure}

To illustrate the dynamical motion of the stars, we plot the coordinate location of one
of the stars in Fig.~\ref{centers} for the two cases. The boxes/circles indicate the
location of the star in the BB/BaB cases at the same interval of an asymptotic observer's
time. Notice that while early on the stars behave roughly in the same manner, as they
get closer the BB case `speeds' up and merges earlier than the BaB. This can be
understood by our analysis indicating the effective Newtonian potential for the BB case
is deeper than that of the BaB case (see the Appendix of~\cite{POLL07}).

\begin{figure}[h]
\begin{center}
\epsfxsize=2.5in
\epsffile[40 18 707 584]{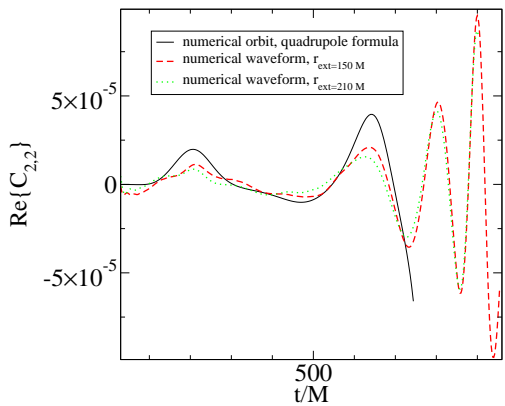}\\
\end{center}
\caption{\textit{Boson/boson pair ($J_z=1.1 M^2$)}. The real part of
the dominant mode $C_{2,2}$ of $M~r~\Psi^4$ for the BB pair, computed from the
numerical waveform at two different extraction surfaces and from the numerical trajectories,
by using the quadrupole formula.}
\label{psi4_bb}
\end{figure}

\begin{figure}[h]
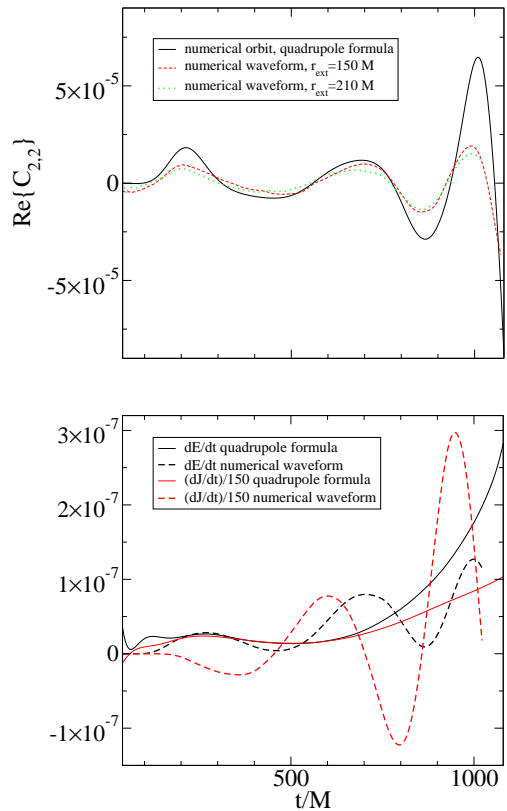

\begin{center}
\epsfxsize=2.5in
\epsffile[40 18 707 584]{c22_bab.eps}\\
\epsfxsize=2.5in
\epsffile[40 59 708 584]{flux_bab.eps}\\
\end{center}
\caption{\textit{Boson/antiboson pair ($J_z=1.1 M^2$)}. The real part of
the dominant mode $C_{2,2}$ of $M~r~\Psi^4$ for the BaB pair at the top, computed from the
numerical waveform at two different extraction surfaces and from the numerical trajectories,
by using the quadrupole formula. The flux of energy and angular momentum are at the bottom,
extracted at the outer extraction surface and compared with the result from the quadrupole formula.
Notice that the flux of angular momentum is at least two orders of magnitude larger than the
flux of energy.}
\label{psi4_bab}
\end{figure}

The dominant mode $l=2,m=2$ of the spin-weighted decomposition of $Mr\Psi_4$ is shown in
Fig.~\ref{psi4_bb} for the BB pair and Fig.~\ref{psi4_bab} for the BaB pair. Notice that
the final speeds are around $v=0.15~c$, that is roughly twice the initial speed. Since the
BB pair orbits faster before merging, the associated gravitational wave output is stronger
than the BaB case. The energy and angular momentum fluxes are shown only for the BaB pair
because the fluxes for the BB pair were very similar.

Following \cite{Buonanno:2006ui}, we can compare three different measures of
the orbital frequencies obtained from the numerical simulations, which can be defined
under some assumptions. The first one is the coordinate frequency
$\omega_c$, obtained from the numerical trajectories of Fig.~\ref{centers}. This quantity
is susceptible to gauge effects because it depends on the dynamical coordinates used during the
evolution. The second one, $\omega_D$, is extracted from the dominant mode of $\Psi_4$, which
can be estimated by
\begin{equation}
   \omega_D = -\frac{1}{m} {\rm Im}\left[ \frac{{\dot{C}}_{l,m}}{C_{l,m}} \right].
\end{equation}
The third one is computed by assuming a Newtonian binary in a circular orbit. In this case, the standard quadrupole 
formula gives the approximate $l=2$ mode of the inspiral waveform 
\begin{equation}\label{C22_NQC}
  C_{2,\pm 2}(t) = 32 \sqrt{\frac{\pi}{5}} \frac{\mu}{M}~[M~\omega(t)]^{8/3}
                   e^{\mp i 2(\phi(t) - \phi_0)}
\end{equation}
where $\mu=M^{(1)} M^{(2)}/M$ is the reduced mass and $\phi(t)$ the accumulated phase of the orbit.
Then, the frequency $\omega_{NQC}$ computed from the Newtonian quadrupole circular orbit approximation
can be computed from (\ref{C22_NQC}) as
\begin{equation}
   M~\omega_{NQC}(t) = \left( \frac{M}{32~\mu}~\sqrt{\frac{5}{\pi}} \left| C_{2,\pm 2}(t) \right| \right)^{3/8}.
\end{equation}

\begin{figure}[h]
\begin{center}
\epsfxsize=7cm
\epsfbox{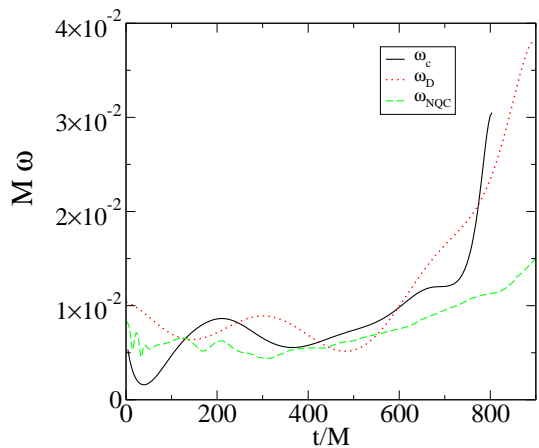}
\end{center}
\caption{\textit{Boson/boson pair ($J_z=1.1 M^2$)}. The orbital frequency computed three different
ways; $\omega_c$ from the numerical simulation, $\omega_D$ from the dominant mode of $\Psi_4$,
and $\omega_{NQC}$ assuming the Newtonian quadrupole circular orbit approximation.}
\label{omega_bb}
\end{figure}

\begin{figure}[h]
\begin{center}
\epsfxsize=7cm
\epsfbox{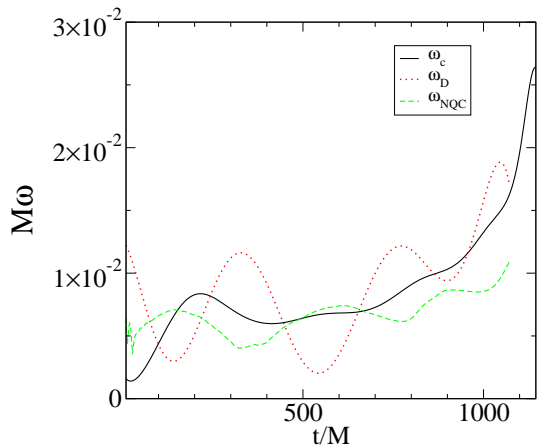}
\end{center}
\caption{\textit{Boson/antiboson pair ($J_z=1.1 M^2$)}. The same quantities as in Fig.~\ref{omega_bb}
but for the boson/antiboson pair. The lower values of the frequencies at the final times, with respect to the
boson/boson pair, is due to the weaker effective interaction between the boson/antiboson stars.}
\label{omega_bab}
\end{figure}

These three measures of the orbital frequency are shown in Fig.~\ref{omega_bb}, where there is a
pretty good agreement between the coordinate frequencies $\omega_c$ and $\omega_D$.
The frequency $\omega_{NQC}$ shows also a remarkable agreement in the first stages, taking into
account that the initial data is only near the Newtonian quasi-circular orbit configuration. 
However, at the last part of the plot, the frequency $\omega_{NQC}$  differs significantly from the others
because its approximation does not consider interactions other than purely gravitational,
namely scalar field interactions.
The same quantities are shown for the boson/antiboson pair in Fig.~\ref{omega_bab}.

\section{Comparisons and conclusion}
\label{conclusions}

In this work we have studied the different phenomenology arising from the addition
of angular momentum in the boson star binaries, going further than in 
our previous work \cite{POLL07}. First, a survey of the possible different behaviours,
--for equal-mass binaries-- was performed by varying the initial angular speed of the stars.
This survey revealed the significantly different outcome that can arise due to the restrictions
single stationary boson stars have as far as their allowed angular momentum.
Next, several cases
were studied in detail for the BB and BaB pairs, paying special attention to configurations 
resulting in orbits. Here the radiation of the system was computed and few salient features
compared with binary black hole simulations. A detailed comparison between binary black holes,
binary neutron star and binary boson stars is undergoing and will be presented elsewhere\cite{AP07}.

It is worth stressing an important consequence of the survey performed. Except for some special cases, 
almost all initially ``bounded'' configurations (as dictated by a Newtonian analysis) 
give rise to an evolution describing either the formation of a black hole or the dispersal of the
scalar field.
Thus it would appear quite difficult to form a single, rotating boson star from
the collision of high mass (and high compaction ratio) boson stars.
Of course, one could try forming rotating boson stars with lower mass boson stars, 
or with other forms for the interaction potential (e.g. terms quartic in the field modulus).
However, our suspicion is that such formation, if it occurs at all,
will turn out to be very much non-generic.

The different particular cases studied showed rich features. The low angular momentum
cases end up with a dispersed scalar field and a low-mass remnant object at the center.
Even more interesting was the realization that in the BaB case, a closely rotating pair
consisting of a boson/antiboson is induced.

For the large angular momentum case, both configurations (BB and BaB) evolve to
a spinning black hole after dispersing/radiating part of its mass and angular momentum.
We computed the gravitational
waveforms produced by the merger for both cases and compared them with the waveform computed
using the quadrupole formula. Interestingly, while the stars are quite close
initially and only perform about an orbit before merging, the computed
waveforms show significant agreement with those obtained from the quadrupole formula. 
Moreover, even in the more violent merger of the BB pair, there is a smooth transition
between the spiral and the merger, as has already been observed in binary black hole mergers.
Hence this system represents
another example of compact binary coalescence for which the nonlinearities of the  merger are
barely manifest in the gravitational waveforms.


\acknowledgments
We would like to thank I.~Olabarrieta, M.~Anderson, M.~Choptuik, D.~Garfinkle, E.~Hirschmann, D.~Neilsen, 
F.~Pretorius, and J.~Pullin  for helpful discussions.
This work was supported in part by NSF grants
PHY0326311 and PHY0554793  to Louisiana State University and PHY-0325224 to Long Island University.
The simulations described here were performed on
local clusters in the Dept. of Physics \& Astronomy at LSU and the
Dept. of Physics at LIU as well as on Teragrid resources provided by
SDSC under allocation award PHY-040027.
L.L. thanks the University of Cordoba for hospitality
where parts of this work were
completed.

\appendix


\bibliographystyle{prsty}

\end{document}